\documentstyle[12pt,epsfig]{article}

\begin{document}
\begin{center}
%{\large{\bf Evidence for Directed Percolation Universality in
%Deterministic Models of Spatiotemporal Intermittency}}\\
  {\large{\bf Evidence for directed percolation universality at the
      onset of spatiotemporal intermittency in coupled circle maps}}\\

\vspace{1cm}

T.M. Janaki$^{a,1}$, Sudeshna Sinha$^{a,2}$ and Neelima
Gupte$^{b,3}$\\ $^a${\it Institute of Mathematical Sciences, Taramani,
Chennai 600 113, India}\footnote{janaki@imsc.ernet.in}$^,
$\footnote{sudeshna@imsc.ernet.in} \\$^b$ {\it Department of Physics,
IIT Madras, Chennai 600036,
India}\footnote{gupte@chaos.iitm.ernet.in}
\end{center}
\begin{abstract}

  We consider a lattice of coupled circle maps, a model arising
  naturally in descriptions of solid state phenomena such as Josephson
  junction arrays. We find that the onset of spatiotemporal
  intermittency (STI) in this system is analogous to directed
  percolation (DP), with the transition being to an unique
  absorbing state for low nonlinearities, and to weakly chaotic
  absorbing states for high nonlinearities. We find that the complete
  set of static exponents and spreading exponents at all critical
  points match those of DP very convincingly. Further, hyperscaling
  relations are fulfilled, leading to independent controls and
  consistency checks of the values of all the critical exponents.
  These results lend strong support to the conjecture that the onset
  of STI in deterministic models belongs to the DP universality class.

\end{abstract}

\newpage

\section{Introduction}
Spatiotemporal intermittency (STI) in coupled map lattices (CML) has
been extensively studied in varied contexts, especially as it is the
precursor of fully developed spatiotemporal chaos in extended
dynamical systems \cite{Kaneko}. There are two types of motion seen in
systems exhibiting STI: laminar and turbulent. The laminar region is
characterized by periodic or even weakly chaotic dynamics, while no
spatiotemporally regular structure can be seen in the turbulent
regime. A laminar or `inactive' site becomes turbulent or `active' at
a particular time only if at least one of its neighbours was turbulent
at an earlier time, i.e., there is no spontaneous creation of
turbulent sites. Hence a turbulent site can either relax spontaneously
to its laminar state or contaminate its neighbours \cite{type1}. This
feature is analogous to directed percolation (DP).  Also, once all the
sites relax spontaneously to its laminar state, the system gets
trapped in this state for all time. Hence, the laminar state is
``absorbing'' in STI. The existence of the absorbing states led to the
conjecture by Pomeau that STI in deterministic models also belongs to
the DP universality class \cite{pomeau}. To be more precise, if the
propagation rate of turbulence is below a certain threshold, the
turbulent states die out and the system remains in the laminar state
for all time (laminar phase/inactive phase). On the other hand, on
exceeding this threshold, the turbulent states start ``percolating''
in spacetime (turbulent/inactive phase).

While there is substantial evidence of DP universality in stochastic
models exhibiting a continuous transition to an absorbing state
\cite{hinrich, Jens}, it is of considerable interest to examine the
robustness of DP critical behaviour in systems with {\em completely
  deterministic evolution rules}. To this end, Chate and Manneville
introduced a simple CML \cite{chate} exhibiting STI and possessing
infinitely many absorbing states. Surprisingly, it was found that not
only were the exponents governing the onset of STI different from
those of DP, they were also non-universal in nature \cite{chate}. This
non-universal behaviour was considered to be due to the existence of
travelling solitary excitations (solitons) with long life times in
this model.  It was argued that the existence of these solitons, which
spoiled the DP nature of the universality class, was an artefact of the
synchronous updating rule and asynchronous updates were tried with
this model. This asynchronous update destroyed the solitons and then
the static exponents were found to be consistent with DP \cite{bohr1}.
Later it was also observed that even finite lifetime solitons could
completely change the nature of transition in a weak soliton region
\cite{bohr2}. This indicates that it is {\em non-trivial to map
  deterministic dynamics to stochastic behaviour} as various
spatiotemporal structures (such as these solitons) may introduce long
range correlations thereby ruining the analogy. Hence it is of
considerable interest to find CMLs with regimes where there are no
additional special spatiotemporal structures, which can be used as
clean testbeds for checking the validity of the DP universality class.

Since the state variables of CMLs can often be identified with
physical quantities (such as voltages, currents, pressures,
temperatures, concentrations or velocities) in fairly realistic
situations, it is conceivable that such models may suggest various
experimental possibilities for observing DP, which still remains an
outstanding problem \cite{hinrich}. Further STI is a common phenomena
of many extended systems, and is seen for instance in experiments on
convection \cite{convect} and in the ``printers instability''
\cite{print}. So if the onset of STI does in fact exhibit DP
universality, then it could lead to promising candidates for observing
DP in real phenomena. And indeed a recent experiment by Rupp {\em et
  al} \cite{ferrofluid} finds agreement with DP exponents at the
transition to STI in a 1-dimensional system of ferrofluid spikes
driven by an external oscillating magnetic field.

Now, in the last decade only the Chate-Manneville CML and its variants
\cite{bohr2} have been studied in this context. Numerical evidence
from more varied sources is required, especially in the absence of
analytical results, in order to settle the question of DP universality
in transitions occuring in deterministic systems. Thus, it is of considerable
interest to be able to find examples of this correspondence in systems
quite distinct from the Chate-Manneville class, and with qualitatively
different absorbing states, that would lend credence to the Pomeau
conjecture. This work provides one such example.

In this paper, we consider the coupled circle map lattice
\cite{nandini} which has been used to model mode-locking behaviour of
the type seen in coupled oscillator systems and in diverse
experimental systems such as Charge Density Waves and
Josephson-Junction arrays \cite{joseph}. We find that this system has
regimes which show spatio-temporal intermittency, with both unique and
weakly chaotic laminar regions, at different values of the
nonlinearity parameter. Importantly, the solitons which spoilt the DP
behaviour in the earlier studies are completely absent here.  Thus we
have at hand a CML without any potentially problematic coherent
spatiotemporal structures, showing the onset of spatiotemporal
intermittency very cleanly. This CML can then serve as a good testbed
for checking the validity of the DP universality class for both unique
and fluctuating absorbing states.

We will now present results from this system strongly indicating that
the complete set of static exponents characterising the transition to
STI are completely consistent with DP, both for unique and weakly
chaotic absorbing states. We will also show that the spreading
exponents (dynamic exponents) for unique as well as fluctuating
absorbing states agree within 3 \% of those obtained in DP. Further we
will demonstrate that the hyperscaling relations in case of the static
as well as the spreading exponents are also satisfied. Thus we will
provide two distinct examples of clean DP universality in transitions
to STI, one of which constitutes the {\em first known example of this
  correspondence in a CML with an unique absorbing state}, and the
other constitutes the {\em first example of this correspondence when
  there exists weakly chaotic absorbing state}.

\section{ Static (Bulk) Exponents}

First, recall the coupled circle map lattice\cite{gauri}:
\begin{equation}
 \theta_{i,t+1}~=~ (1-\epsilon)~f(\theta_i,t) +
 \frac{\epsilon}{2}~(f(\theta_{i-1,t})+f(\theta_{i+1,t})) \ \ \
{\rm mod} \ \ 1 \nonumber
\end{equation}
\noindent where $t$ is the discrete time index, and $i$ is the site
 index: $i=1, \dots L $, with $L$ being the system size. The parameter
 $\epsilon$ gives the strength of the diffusive coupling between site
 $i$ and its two neighbours. The local on-site map is given by
\begin{equation}
f(\theta)~=~ \theta + \omega -\frac{k}{2 \pi} \sin(2 \pi \theta) \nonumber
\end{equation}
where the parameter $k$ gives the nonlinearity. This CML has been
studied extensively with parallel up-dates and has a rich phase
diagram with many types of attractors and strong sensitivity to
initial conditions\cite{gauri,gauri1}. In particular, this system also
has regimes of spatio-temporal intermittency (STI) when evolved
parallely with random initial conditions. Figs.~1 and 2 show
space-time plots of the spatio-temporal intermittency observed in two
different STI regimes. It is clear that no travelling wave
soliton-like structures are seen in this regime.  Hence, it is not
necessary to introduce any asynchronicity here to destroy
``solitonic'' behaviour, as this system is naturally free of such
spatiotemporal excitations in the parameter region studied here.

We shall now study the onset of spatiotemporal intermittency in this
system. Interestingly, as mentioned before, two qualitatively distinct
absorbing regions can be found in this system.

(i) When nonlinearity parameter $k = 1$, there are regions of
$(\epsilon - \omega)$ space where the system goes to the synchronised
spatiotemporal fixed point $\theta^{\star} = \frac{1}{2\pi}
\sin^{-1}(\frac{2\pi \omega}{k})$. This constitutes an {\em unique}
absorbing state (see Fig.~1). We closely scrutinise the critical
behaviour at 2 critical points in this regime: $\omega=0.064,
\epsilon=0.63775$, and $\omega=0.068, \epsilon=0.73277$. These mark
the transition from a laminar phase to STI. The turbulent sites here
are those which are different from $\theta^{\star}$.

(ii) When nonlinearity parameter $k = 3.1$, there are regions of
$(\epsilon - \omega)$ space where sites with any value less than $1/2$
constitute the absorbing states, and sites whose values are greater
than $1/2$ constitute the turbulent states (see Figs.~2 a-b). So now
the absorbing states are {\em inifinitely many}, as also {\em weakly
  chaotic}. We study the critical behaviour at 2 critical points in
this regime: $\omega=0.18, \epsilon=0.701$, and $\omega=0.19,
\epsilon=0.65612$.

As mentioned earlier, we initiate the evolution with random initial
conditions and let the system evolve under parallel updates.  The DP
universality class is characterised by a set of critical exponents
which describe the scaling behaviour of the quantities of physical
interest.  The physical quantities of interest for such systems are
(a) the escape time $\tau$, which is the number of time steps elapsed
before the system reaches its laminar state and (b) the order
parameter, $m(\epsilon, L, t)$, which is the fraction of turbulent
sites in the lattice at time $t$.  From finite-size scaling arguments,
it is expected that $\tau$ depends on $L$ such that
 \[ \tau ( \omega, \epsilon ) =\left \{
\begin{array}{ll} log~L & \mbox{laminar phase}\\ L^z &\mbox{critical
phase}\\ exp~ L^c & \mbox{turbulent phase} \end{array} \right. \]
Here, $c$ is a constant of order unity, and  the critical point is
identified as the set of parameter values at which $\tau$ shows
power-law behaviour with $z$ being the critical exponent.
At $\epsilon_c$ the critical value of the parameter $\epsilon$ (other
parameters being held fixed), the order parameter $m(\epsilon,L,t)$ scales as
\begin{equation}
m \sim (\epsilon-\epsilon_c)^{\beta} ,~~\epsilon \rightarrow \epsilon_c^{+}.
\end{equation}
when the critical line is approached from above. Also, at the critical
$\epsilon_c$, the order parameter is expected to satisfy the scaling
relation
\begin{equation}
m \sim (\epsilon-\epsilon_c)^{\beta} ,~~\epsilon \rightarrow \epsilon_c^{+}.
\end{equation}

We compute the above quantities for our CML averaged over an ensemble
of $10^4$ initial conditions.  The dependence of $\tau$ on $L$ for
different values of $\epsilon$ is shown on a log-log plot in Fig.~3.
Fig.~3 shows this dependence at the parameter values $k=1$, $\omega =
0.068$ (i.e. at parameter values which correspond to a unique
absorbing state). It is clear from the graph that an algebraic
increase can be seen at the value $\epsilon_c= 0.63775$. A similar
analysis was carried out for the parameter values $k = 1$,
$\omega=0.064$, where a unique absorbing state can again be seen and
gave the critical value $\epsilon_c = 0.73277$. Weakly chaotic
absorbing states were seen at the parameter values $k = 3.1$,
$\omega=0.18$, and $\omega=0.19$. Here the critical values of
parameters turned out to be $\epsilon_c = 0.70100$ for the first case
and $\epsilon_c = 0.65612$ for the second case.  The critical exponent
$z$ was estimated at these critical values for all four cases. The
log-log plot of the escape time $\tau$ against the system size $L$ is
shown in Fig.~4. It is clear that the same exponent $z$ is seen for
all four cases and turns out to lie in the range $1.58-1.59$ which is
completely consistent with the DP value for this exponent. (See Table
I).

The order parameter is expected to obey the scaling relation
\begin{equation}
m(\epsilon_c, L,t) \approx  t^{-\beta / \nu  z}
\end{equation}

for $ t << \tau$.  Therefore the log-log plots of $m$ as a function of
time $t$ for various lattice sizes must fall on one line when $ t <<
\tau$ and the power $- \beta / \nu z$ must correspond to the slope of
the graph for these regimes.  The order parameter is plotted as a
function of $t$ on log-log plot in Figs.~5 and 6, where the data in
Fig.~5 is obtained for the parameter values $k = 1$, $\omega=0.068$
and $\epsilon_c=0.63775$ (the unique absorbing state case, and that in
Fig.~6 is obtained for $k = 3.1$, $\omega=0.19$ and
$\epsilon_c=0.65612$ (the case with weakly chaotic absorbing states).
In both cases, the data for $L= 50, 100, 300, 500, 1000$ collapses on
to one line, the slope of which gives $- \beta / \nu z = - 0.16$.

The order parameter of systems which belong to the directed
percolation universality class satisfies the scaling function

\begin{equation}
m(\epsilon_c, L, t) \sim L^{\beta / \nu} g_{ml}(t/L^z). \label{col1}
\end{equation}
at the critical value $\epsilon=\epsilon_c$. We plot the order
parameter for our CML at the critical values above in Figs. 7 and 8.
The data with scaled variables $M = m L^{\frac{\beta}{\nu}}$ and
$T=t/L^z$ fall on one curve for various lattice sizes indicating
dynamical scaling (see Figs.~7 and 8, parameter values are as given in
the figure captions). This further substantiates the claim that the
behaviour of the sine circle map CML falls in the directed percolation
universality class.

The exponent $\nu$ can be extracted independently, by using the
scaling relation
\begin{equation}
 \tau (L,\delta) \approx \phi ^z f(L / \phi),
\end{equation}
where $\phi$ is the correlation length which diverges as $\phi \approx
\delta ^{- \nu}$ and $\delta $ is given by $\epsilon - \epsilon_c$
\cite{Houl}. Therefore, $\nu$ can be obtained by adjusting it's value
till the scaled variables $L \delta^{\nu}$ and $\tau \delta^{\nu z}$
collapse onto a single curve.  Thus, the exponent $\beta$ can be
obtained from equation \ref{col1}.

\noindent  To extract further critical exponents, we obtain the
correlations from the pair correlation function given by:
\begin{equation}
C_j(t)=\frac{1}{L} \sum_{i=1}^L <u_i(t)u_{i+j}>-<u_i(t)>^2
\end{equation}
where the brackets denote the averaging over different
initial conditions. At criticality one expects an algebraic decay of
correlation, i.e.
$$C_j(t) \approx j^{1- \eta^{\prime}}$$
where $\eta^{\prime}$ is the
associated critical exponent.  The log-log plot of the spatial
correlation function at $k = 1$, $\omega=0.064$, $\epsilon_c=0.73277$
can be seen in Fig.~9.  The log-log plot of the correlation function
approaches a straight line with slope $1-\eta^{\prime}$ at large
times. The value of the exponent $\eta$turns out to be $0.302$, which
is consistent with the directed percolation value. The values of the
exponent at the other critical set of parameter values are listed in
Table I. Directed percolation like behaviour is observed for the
entire set.

Thus we have obtained the {\em complete set of static (bulk)
  exponents}, namely $z,~\beta ,~\nu , ~\eta^{\prime}$, that
characterize the DP class (~see Table I~). Clearly the values of the
exponents obtained for the coupled circle map lattice are in excellent
agreement with the DP values for both unique as well as the weakly
chaotic absorbing states.

\begin{table}
\begin{center}
\begin{tabular}{|c|c|c|c|c|c|c|}
\hline\\
$k$ & $\omega$ & $\epsilon$& $z$ &$\beta$ & $\nu$ & $\eta^{\prime}$\\
\hline\\
$1$ & $0.068$ & $ 0.63775$ & $ 1.580$ & $0.28$ & $1.10$ & $1.49$  \\
$1$ & $0.064$ & $0.73277$ & $1.591$ & $0.28$ & $1.10$ & $1.50$ \\
$3.1$ & $0.18$ & $0.70100$ & $1.597$ & $0.26$ & $1.12$ & $1.50$\\
$3.1$ & $0.19$ & $0.65612$ & $1.591$ & $0.28$ & $1.10$ & $1.49$\\
$ DP$ & $~~$ & $~~$ & $1.58$ & $0.28$ & $1.10$ & $1.51$\\
\hline
\end{tabular}
\end{center}
\caption[]{ Critical static exponents of the synchronously updated
coupled circle map lattice for 4 critical points. The first two critical
points correspond to transitions to an unique absorbing case, while
the third and fourth points correspond to weakly chaotic absorbing states.
The last row shows the corresponding exponents of directed percolation.}
\end{table}

\noindent The exponents also satisfy the hyperscaling relation,
 $2 \beta / \nu = d - 2 + \eta^{\prime}$ , where $d=1$.

\section{ Spreading Exponents}

In the previous section we obtained the static exponents, also known
as the bulk exponents for our CML, and found that they were in good
agreement with those of DP. We shall now compute a set of dynamical
exponents, called the spreading exponents, from the temporal evolution
of a nearly absorbing system with a localized disturbance i.e. with
only few contiguous active (turbulent) sites in an otherwise absorbing
state.  The quantities of interest are, the time dependence of $N(t)$,
the number of active sites at time $t$ averaged over all runs, $P(t)$,
the survival probability, or the fraction of initial conditions which
show a non-zero number of active sites (or a propagating disturbance)
at time $t$ and $R^2(t)$, the mean squared deviation from the origin
of the turbulent activity averaged over surviving runs alone. The
spreading exponents are obtained from the time dependence of these
quantities which show scaling behaviour at criticality. At
criticality, we have,
$$N(t) \approx t^{\eta},~~P(t) \approx t^{- \delta},~~R^2(t) \approx
t^{z_s}.$$
Also, $\delta=\beta / \nu z$. We shall now compute these quantities
for our system and compare them with those of DP.

For $k=1$, we have already seen that the absorbing state is the
synchronised spatiotemporal fixed point. However, when we start with
a lattice with only a single active site and all other sites at the
spatiotemporal fixed point, we see that the sytem goes to its
absorbing state in about $10$ time steps of evolution. Further the
symmetric diffusive coupling in the CML indicates that the temporal
spreading will be symmetric about the single active site, which is not
desirable. Therefore, for good statistics and to counter the symmetric
spreading, we need at least two contiguous active sites (see Fig.~10).
So in our calculations, we have started with two or more contiguous
active sites, while the background is fixed at $\theta^{\star}$. We
find that the full set of spreading exponents obtained thus (see
Figs.~12-14) agree within 3 \% of the DP values (see Table II).

For $k=3.1$ there is no unique absorbing configuration as above, and
any initial configuration with values less than $1/2$ constitutes an
absorbing background. We find that we can now obtain reasonable
asymmetric spreading (and consequently the exponents) with just a
single active seed \cite{singleseed}, and these exponents are the same
as those obtained from spreading from two or three contiguous active
sites (see Fig.~11). Again the complete set of spreading exponents
agree with those of DP within 3 \% at all critical points (see Table
II and Figs.~12-14).

\begin{table}
\begin{center}
\begin{tabular}{|c|c|c|c|c|c|}
\hline\\
$k$ & $\omega$ & $\epsilon$& $\eta$ &$\delta$ & $z_s$ \\
\hline\\
$1$ & $0.068$ & $ 0.63775$ & $0.292 $ & $0.153$ & $1.243$ \\
$1$ & $0.064$ & $0.73277$ & $0.302$ & $0.158$ & $1.259$\\
$3.1$ & $0.18$ & $0.70100$ & $0.310$ & $0.157$ & $1.272$ \\
$3.1$ & $0.19$ & $0.65612$ & $0.308$ & $0.156$ & $1.251$ \\
$ DP$ & $~~$ & $~~$ & $0.313$ & $0.159$ & $1.26$ \\
\hline
\end{tabular}
\end{center}
\caption[]{ Spreading exponents of the synchronously updated
coupled circle map lattice for 4 critical points. Two active seeds in
an absorbing configuration is used as initial condition. For the first
2 critical points there exists an unique absorbing state, while for
the third and fourth points one can have many different absorbing
states and consequently many different initial absorbing backgrounds.
However we notice that the exponents obtained are quite the same for
different initial preparations and thus appears universal for 2 active
seeds.  The last row shows the corresponding exponents of directed
percolation.}
\end{table}

\section{Conclusions}

The evaluation of the complete set of static and spreading exponents
at the onset of spatiotemporal intermittency in coupled circle map
lattices, shows that this transition clearly falls in the universality
class of directed percolation. All the critical characteristics of
directed percolation, such as hyperscaling, are fulfilled, leading to
independent controls and consistency checks of the values of all the
critical exponents. DP exponents are seen at low values of
nonlinearity for a unique absorbing state and at high values of
nonlinearity for weakly chaotic absorbing states. It is not necessary
to introduce asynchronicity or an extra dimension to tune out
solitonic behaviour since no solitonic behaviour is seen in this
model. Thus this model constitutes a very clean system where DP
exponents are very naturally seen. Model studies such as these,
showing the clean correspondence between the onset of STI and directed
percolation, could then lead to promising new candidates for observing
DP in real phenomena.

\pagestyle{empty}

\begin{figure}
\begin{center}
\mbox{\epsfig{file=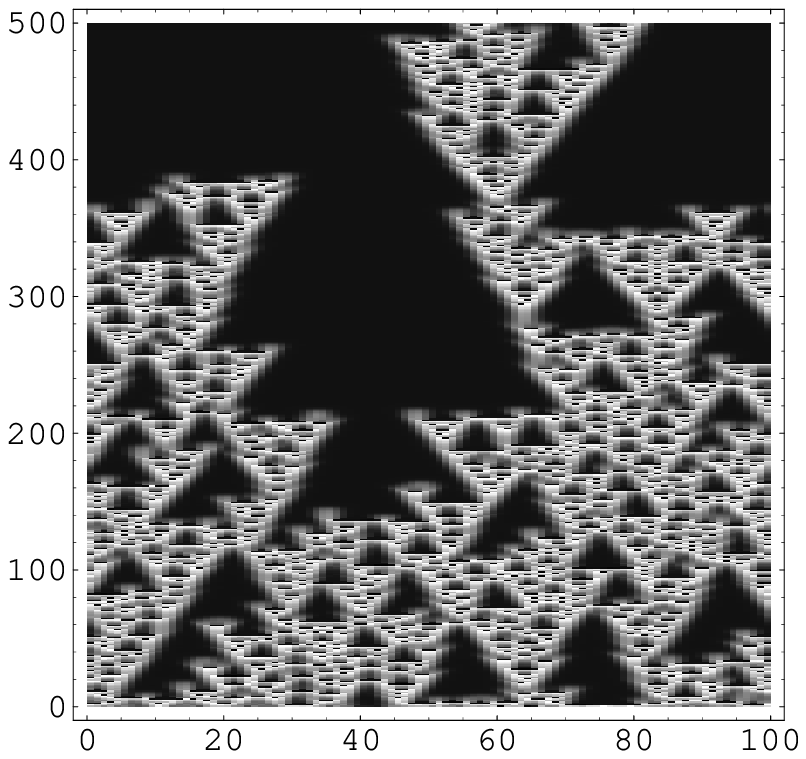,height=10truecm}}
\caption{ STI in a synchronously updated coupled circle map lattice of
  size $L=100$ with parameters $k=1$, $\omega=0.068$, $ \epsilon_c =
  0.73277$. The horizontal axis is the site index $i = 1, \dots L$
  and the vertical axis is discrete time $t$. The absorbing region
  (black) has sites at the spatiotemporal fixed point of the system.}
\end{center}
\end{figure}

\begin{figure}
\begin{center}
\mbox{\epsfig{file=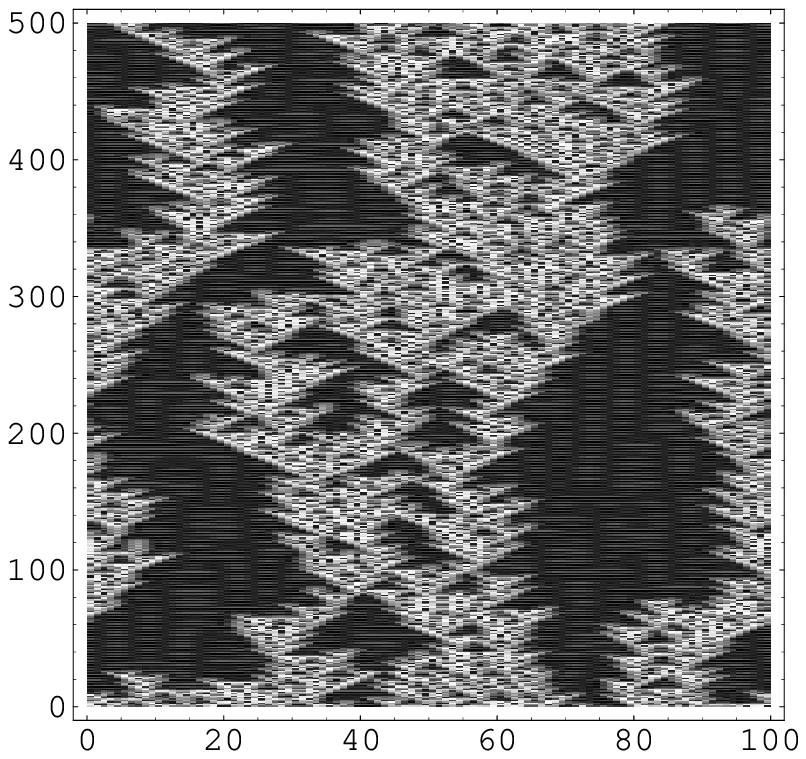,height=9truecm}}
\mbox{\epsfig{file=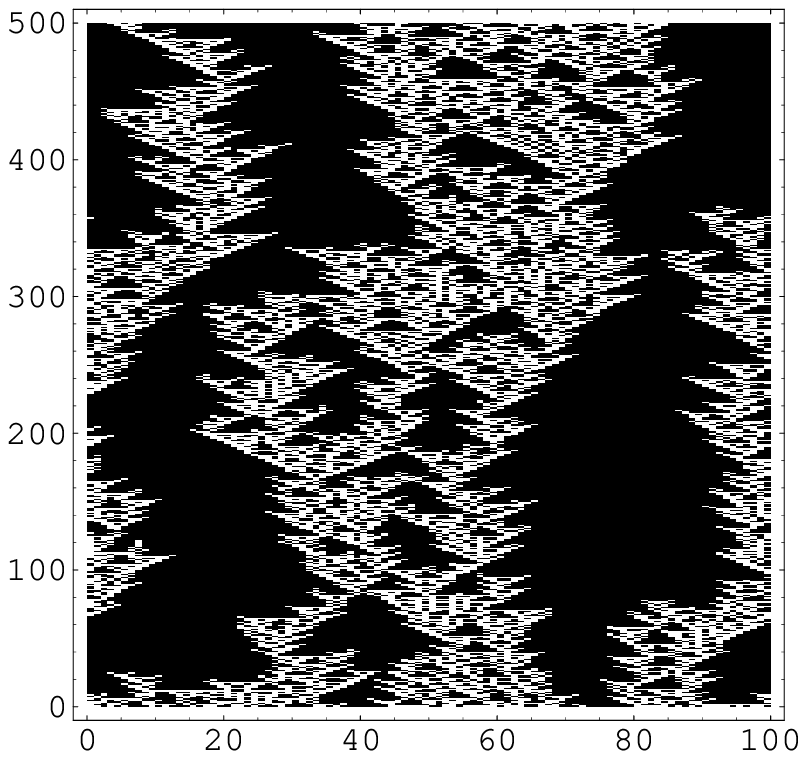,height=9truecm}}
\caption{ STI in a synchronously updated coupled circle map lattice of
  size $L=100$ with parameters $k=3.1$, $\omega=0.18$, $ \epsilon_c =
  0.701$, where the absorbing region has sites below $1/2$ and is not
  unique. The horizontal axis is the site index $i = 1, \dots L$ and
  the vertical axis is discrete time $t$. The top figure is obtained
  from a density plot of the actual $\theta$ values (the absorbing
  regions appear dark). The figure below is a coarse grained plot of
  the above (black is an absorbing site and white is an active one).}
\end{center}
\end{figure}

\begin{figure}[htb]
\begin{center}
\mbox{\psfig{file=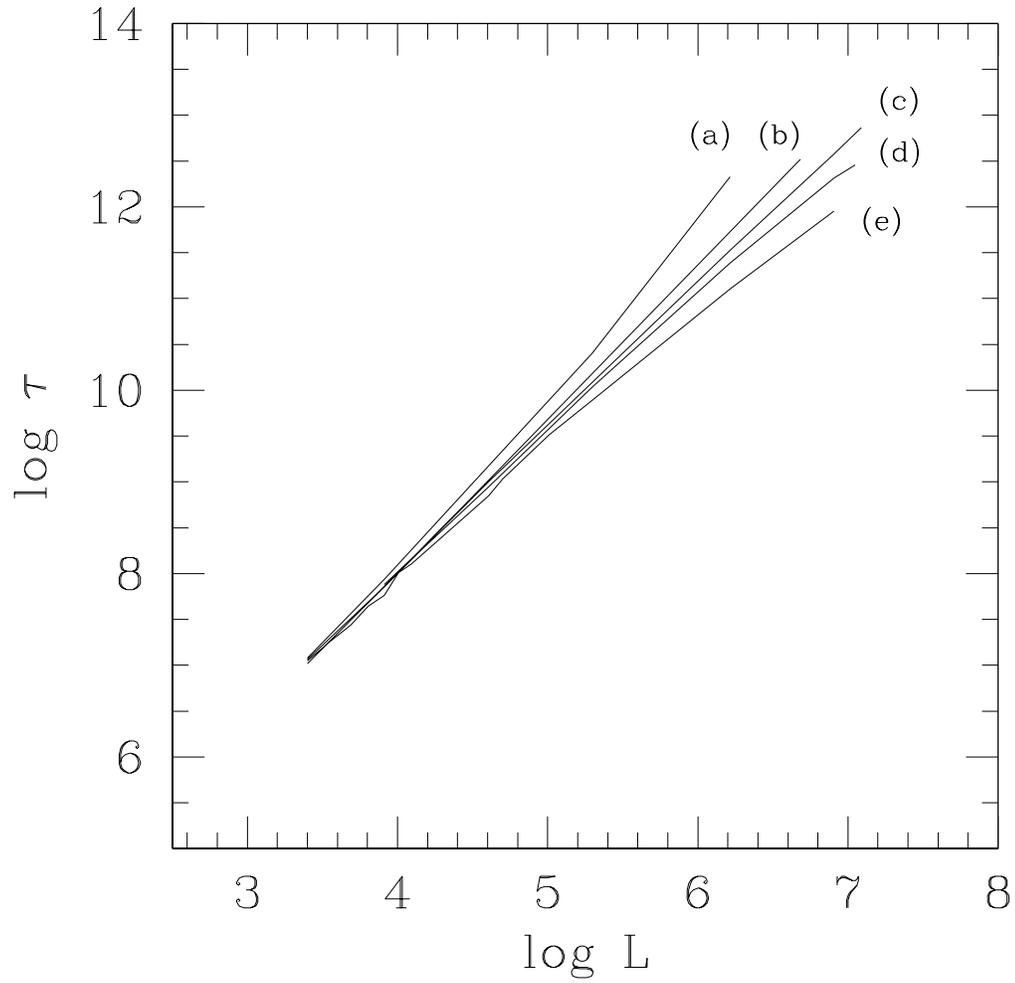,height=15truecm}}
\caption{ Log-log plot (base $e$) of escape time $\tau$ versus lattice
  size $L$ for parameters $k=1$, $\omega = 0.068$ and (a) $\epsilon =
  0.639$, (b)$\epsilon = 0.638$, (c) $\epsilon_c = 0.63775$, (d)
  $\epsilon = 0.637$ and (e) $\epsilon = 0.635$. }
\end{center}
\end{figure}

\begin{figure}
\begin{center}
\mbox{\psfig{file=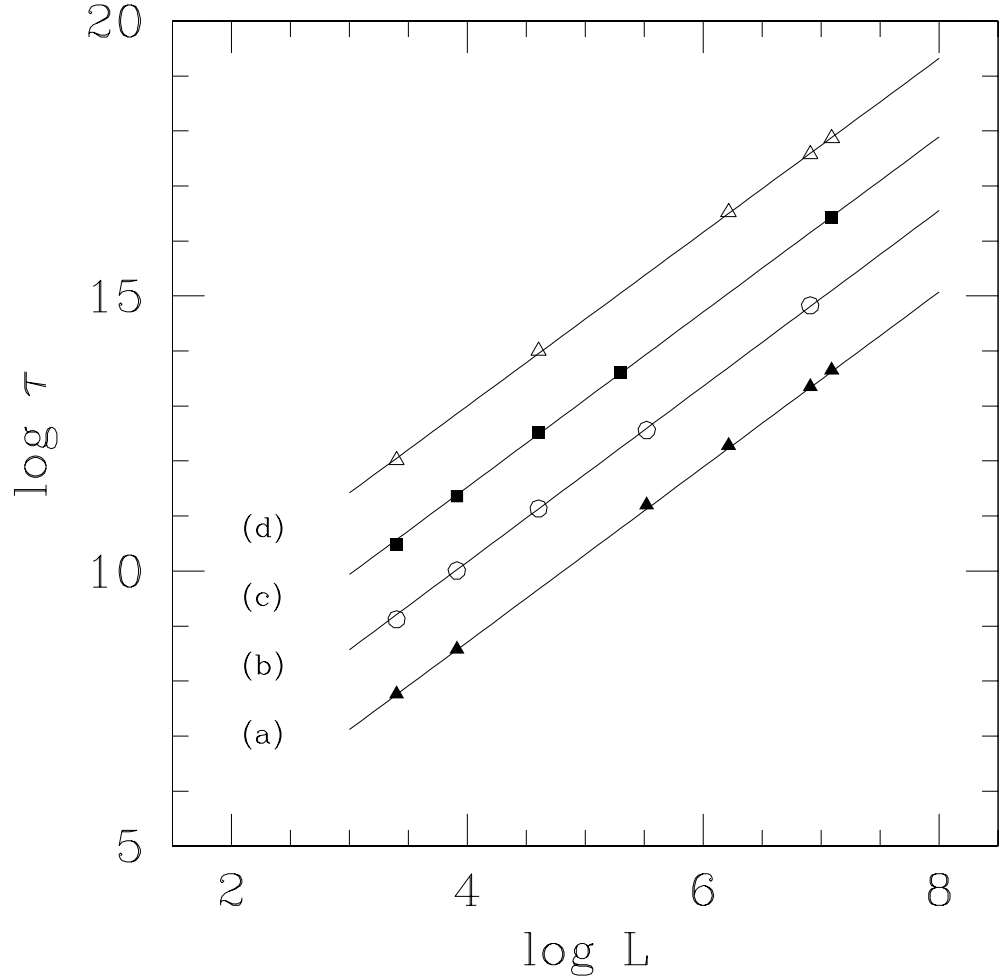,width=15truecm}}
\caption{ Log-log plot (base $e$) of escape time $\tau$ versus lattice
  size $L$ for all 4 critical points: (a) $k = 1$, $\omega=0.068$,
  $\epsilon_c = 0.63775$ and (b) $k = 1$, $\omega=0.064$, $\epsilon_c
  = 0.73277$ (when there exists an unique absorbing state); (c) $k =
  3.1$, $\omega=0.18$, $\epsilon_c = 0.70100$ and (d) $k = 3.1$,
  $\omega=0.19$, $\epsilon_c = 0.65612$ (with fluctuating absorbing
  states). Table I gives the exponent $z$ of the power law fits for
  the different critical points.}
\end{center}
\end{figure}

\begin{figure}[htb]
\begin{center}
\mbox{\psfig{file=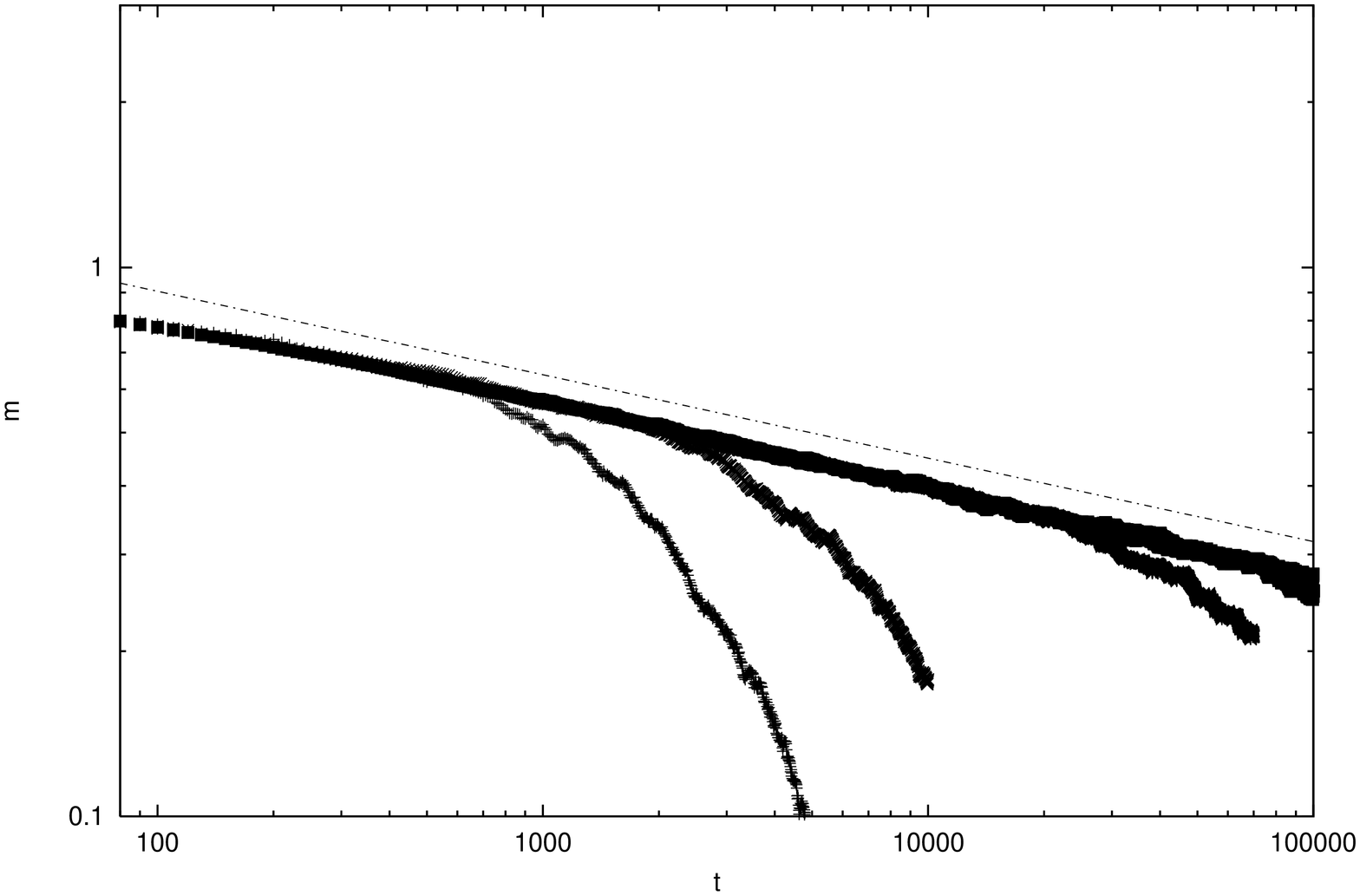,height=15truecm,width=10truecm,angle=270}}
\caption{ Log-log plot of order parameter $m (t)$ versus $ t$
  for different lattice sizes at the critical point: $k = 1$,
  $\omega=0.068$ and $\epsilon_c=0.63775$ (when there exists an unique
  absorbing state).  For $t << \tau$, the data collapses for $L= 50,
  100, 300, 500, 1000$ on to one line, the slope of which gives $-
  \beta / \nu z = - 0.16$}
\end{center}
\end{figure}

\begin{figure}[htb]
\begin{center}
\mbox{\psfig{file=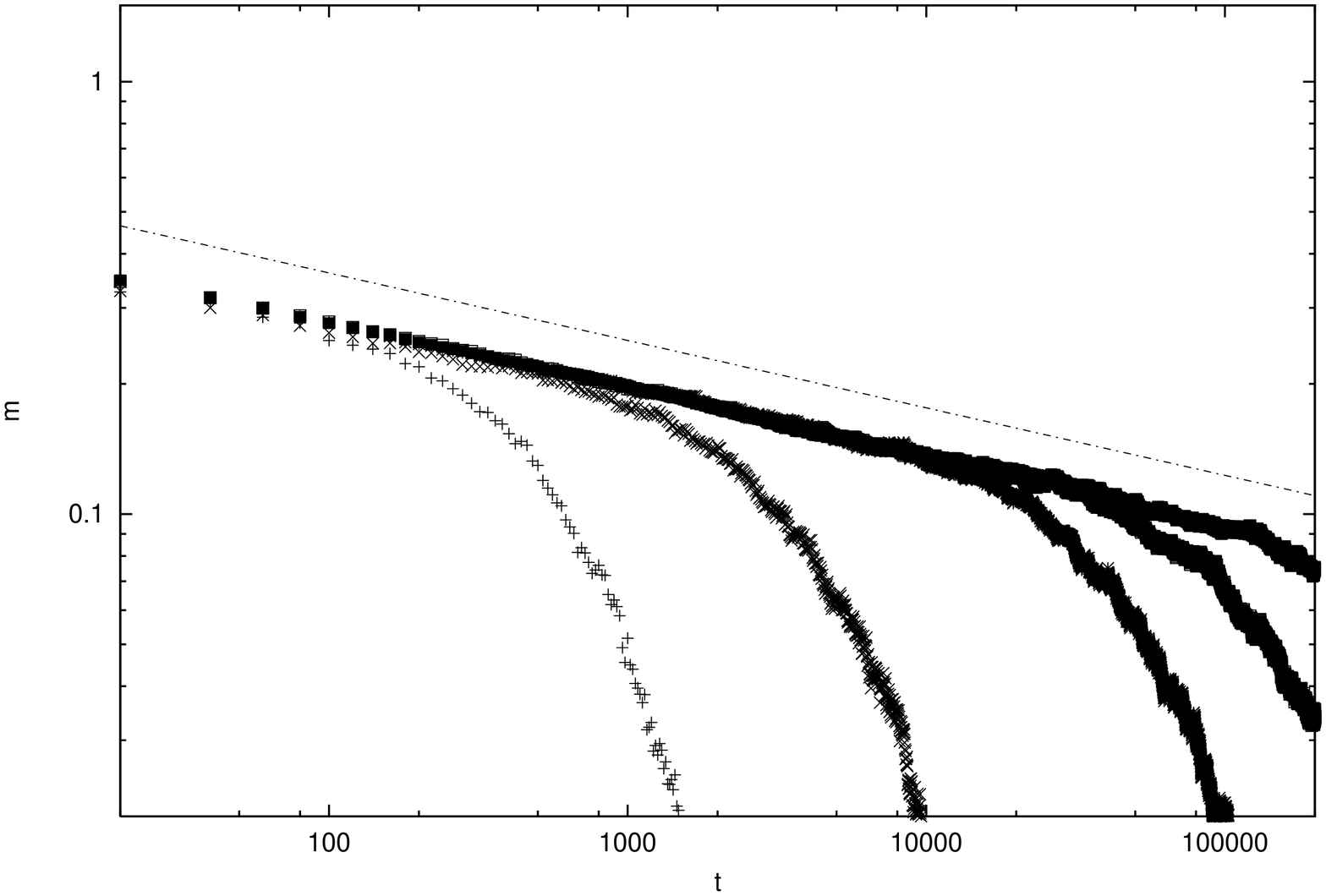,height=15truecm,width=10truecm,angle=270}}
\caption{ Log-log plot of order parameter $m (t)$ versus $ t$
  for different lattice sizes at the critical point: $k = 3.1$,
  $\omega=0.19$ and $\epsilon_c=0.65612$ (a case with inifinitely many
  absorbing states). For $t << \tau$, the data collapses for $L= 50,
  100, 300, 500, 1000$ on to one line, the slope of which gives $-
  \beta / \nu z = - 0.16$.}
\end{center}
\end{figure}

\begin{figure}[htb]
\begin{center}
\mbox{\psfig{file=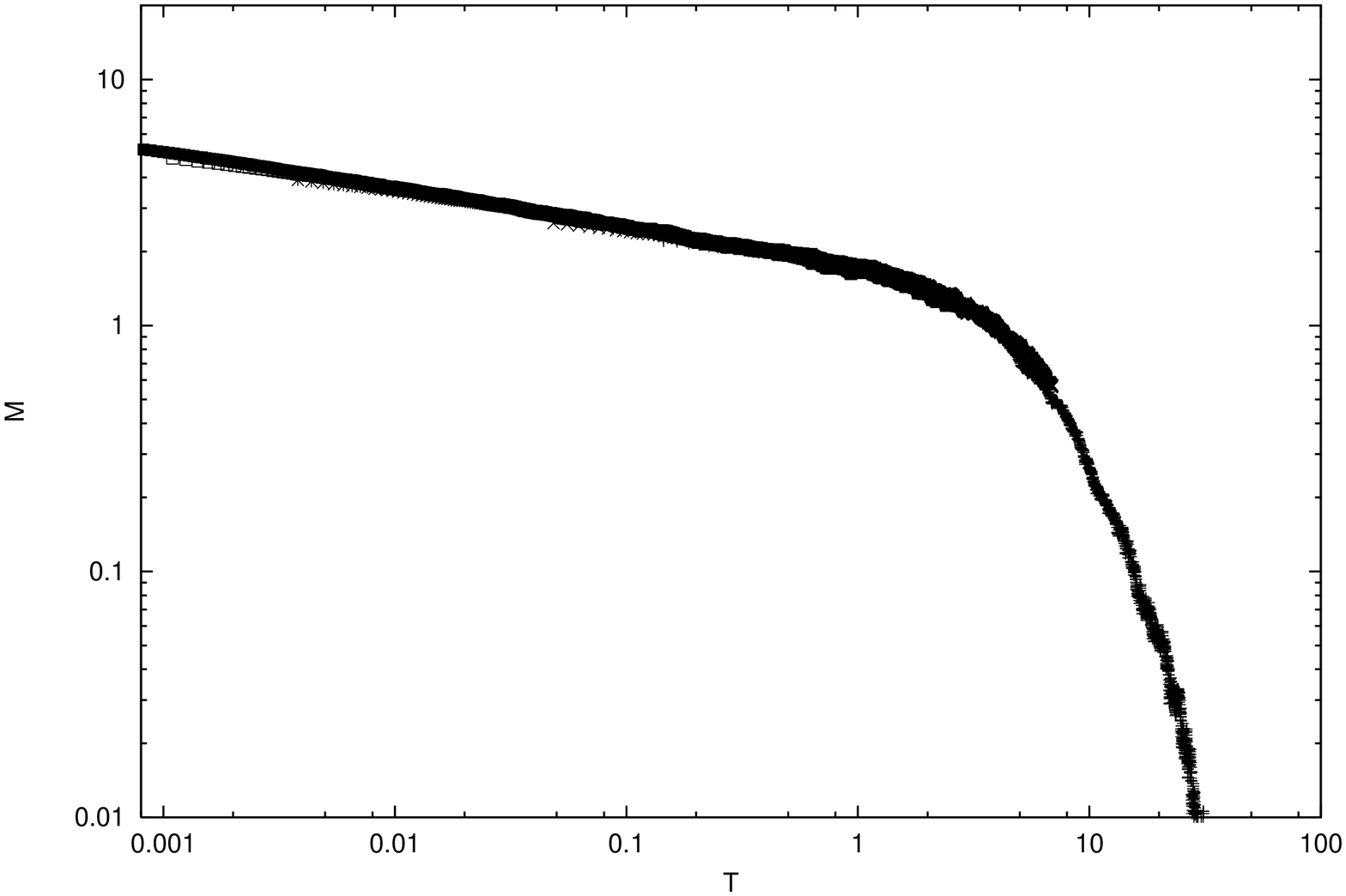,height=15truecm,angle =-90}}
\caption{ Log-log plot of $M = m L^{\beta/\nu}$ vs $T = t/ L^Z$ at the
  critical point: $k = 1$, $\omega=0.068$ and $\epsilon_c=0.63775$
  (when there exists an unique absorbing state). This rescaling of the
  order parameter according to Eqn.~\ref{col1} yields independent
  estimates for $\beta/\nu$ and $z$. The data for system sizes $L$
  ranging from $2^4$ to $2^{10}$ collapses onto one curve.}
\end{center}
\end{figure}

\begin{figure}[htb]
\begin{center}
\mbox{\psfig{file=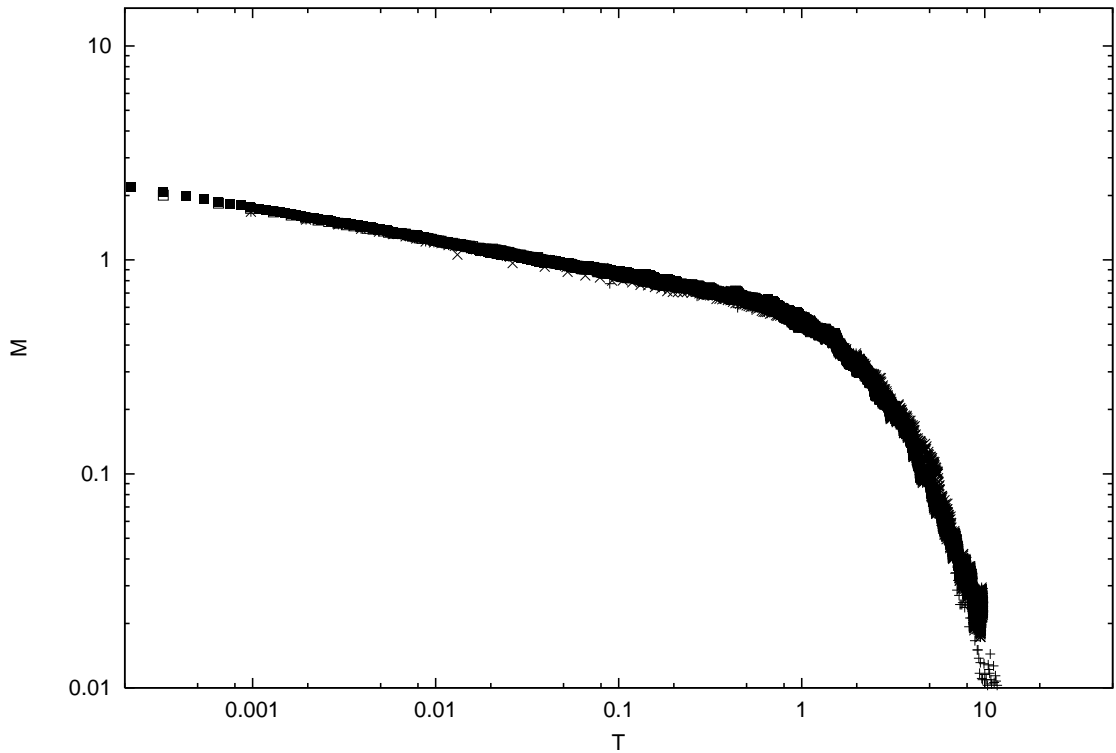,height=15truecm,angle=-90}}
\caption{ Log-log plot of $M = m L^{\beta/\nu}$ vs $T = t/ L^Z$ at the
  critical point: $k = 3.1$, $\omega=0.19$ and $\epsilon_c=0.65612$
  (when there are weakly chaotic absorbing states). This rescaling of
  the order parameter according to Eqn.~\ref{col1} yields independent
  estimates for $\beta/\nu$ and $z$. The data for system sizes $L$
  ranging from $2^4$ to $2^{10}$ collapses onto one curve.}
\end{center}
\end{figure}

\begin{figure}[htb]
\begin{center}
\mbox{\psfig{file=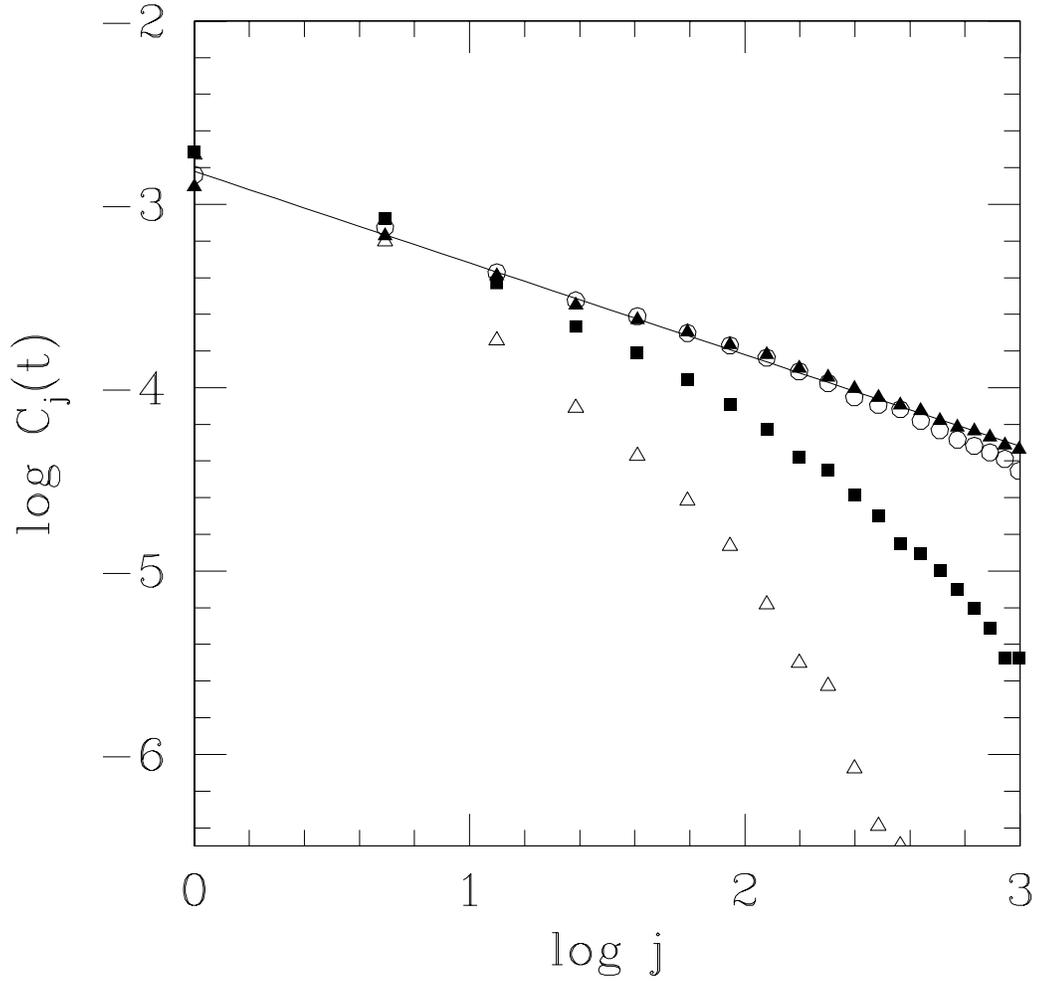,height=15truecm}}
\caption{ Log-log plot (base $e$) of the spatial correlation function
  $C_j (t)$ vs $j$ at various times $t$ ($= 100, 300, 500, 1000$) at
  the critical point: $k = 1$, $\omega=0.064$, $\epsilon_c=0.73277$.
  $C_j (t)$ approaches a straight line with slope $1-\eta^{\prime}$
  for large times indicating an algebraic decay. }
\end{center}
\end{figure}

\begin{figure}
\begin{center}
\mbox{\epsfig{file=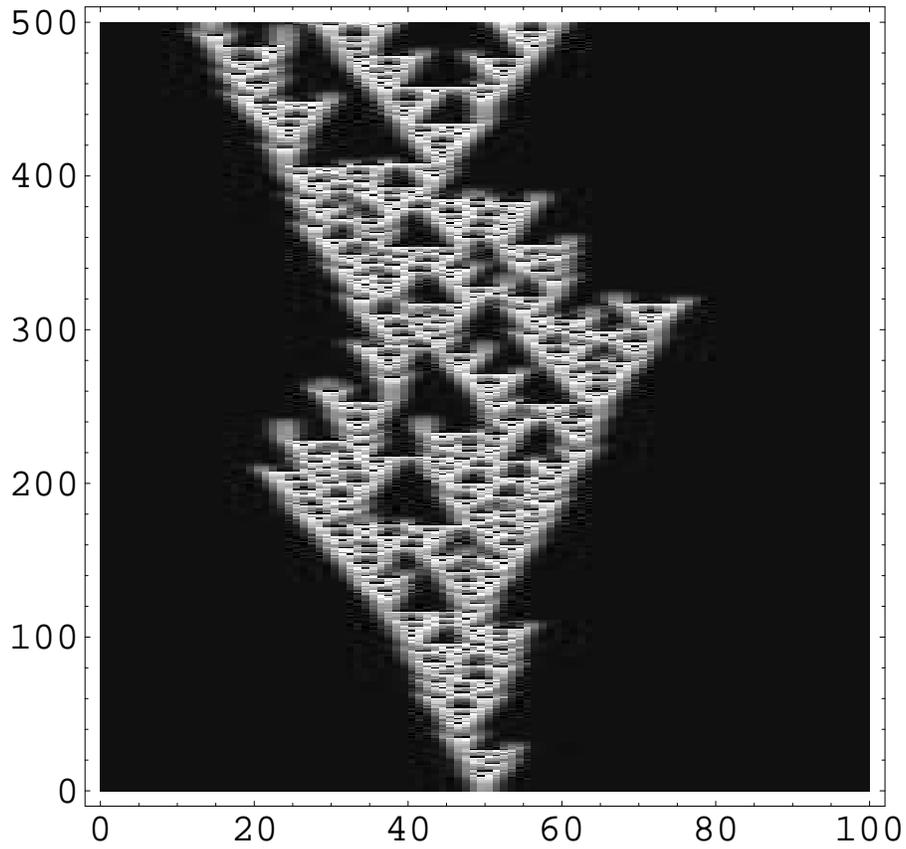,height=12truecm}}
\caption{ Spreading of 2 active seeds in an otherwise absorbing lattice
  of size $L=100$ at critical point $k=1$, $\omega=0.068$, $
  \epsilon_c = 0.73277$. The horizontal axis is the site index $i = 1,
  \dots L$ and the vertical axis is discrete time $t$. The absorbing
  region (black) is unique here, with all sites at the spatiotemporal
  fixed point of the system.}
\end{center}
\end{figure}

\begin{figure}
\begin{center}
\mbox{\epsfig{file=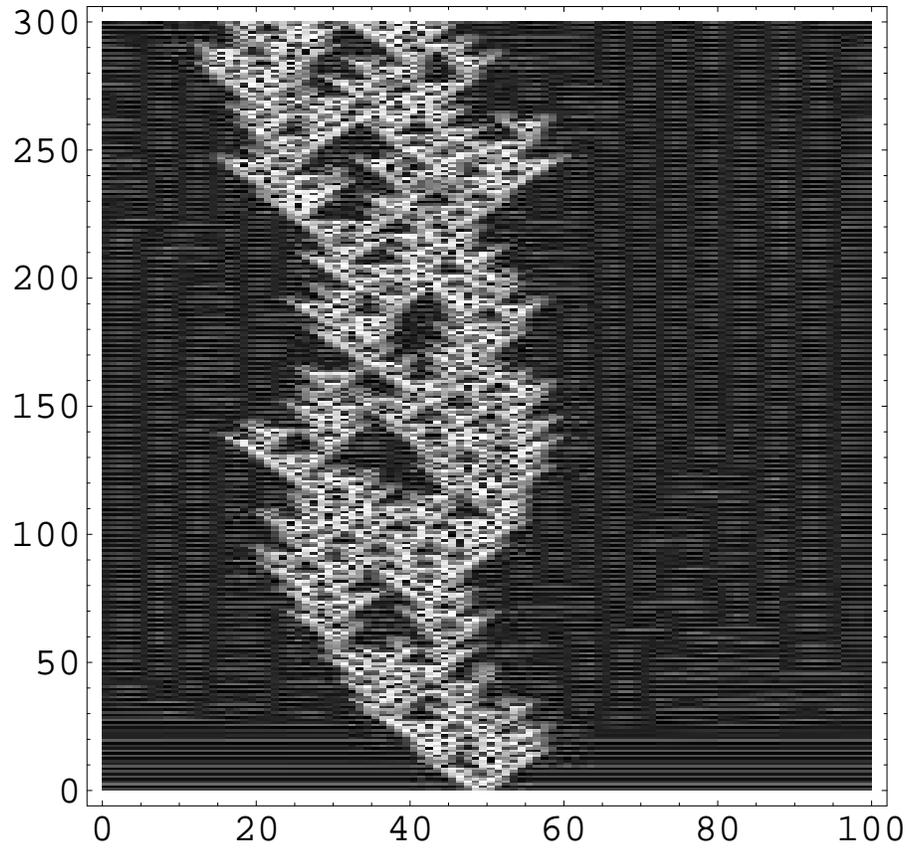,height=12truecm}}
\caption{ Spreading of 2 active seeds in an otherwise absorbing lattice
  of size $L=100$ at critical point $k=3.1$, $\omega=0.18$, $
  \epsilon_c = 0.701$. The horizontal axis is the site index $i = 1,
  \dots L$ and the vertical axis is discrete time $t$. The absorbing
  region (dark) has all sites below $1/2$ and is not unique.}
\end{center}
\end{figure}

\begin{figure}[htb]
\begin{center}
\mbox{\psfig{file=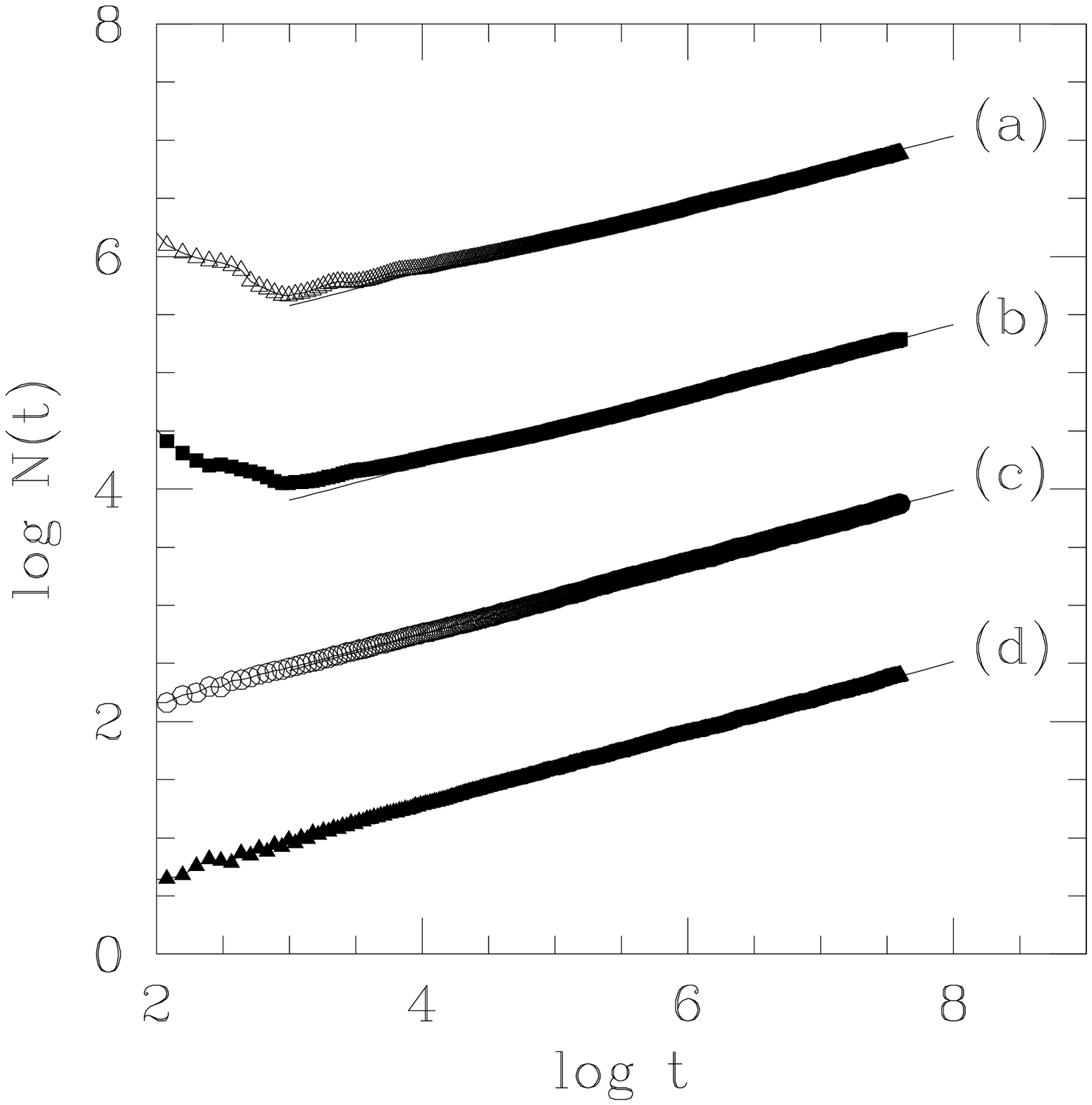,height=15truecm}}
\caption{ Log-log plot (base $e$) of $N (t)$ vs $t$ for all 4 critical
  points: (a) $k = 1$, $\omega=0.068$, $\epsilon_c = 0.63775$; (b) $k
  = 1$, $\omega=0.064$, $\epsilon_c = 0.73277$; (c) $k = 3.1$,
  $\omega=0.18$, $\epsilon_c = 0.70100$ and (d) $k = 3.1$,
  $\omega=0.19$, $\epsilon_c = 0.65612$. Table II gives the exponent
  $\eta$ of the power law fits for the different critical points.}
\end{center}
\end{figure}

\begin{figure}[htb]
\begin{center}
\mbox{\psfig{file=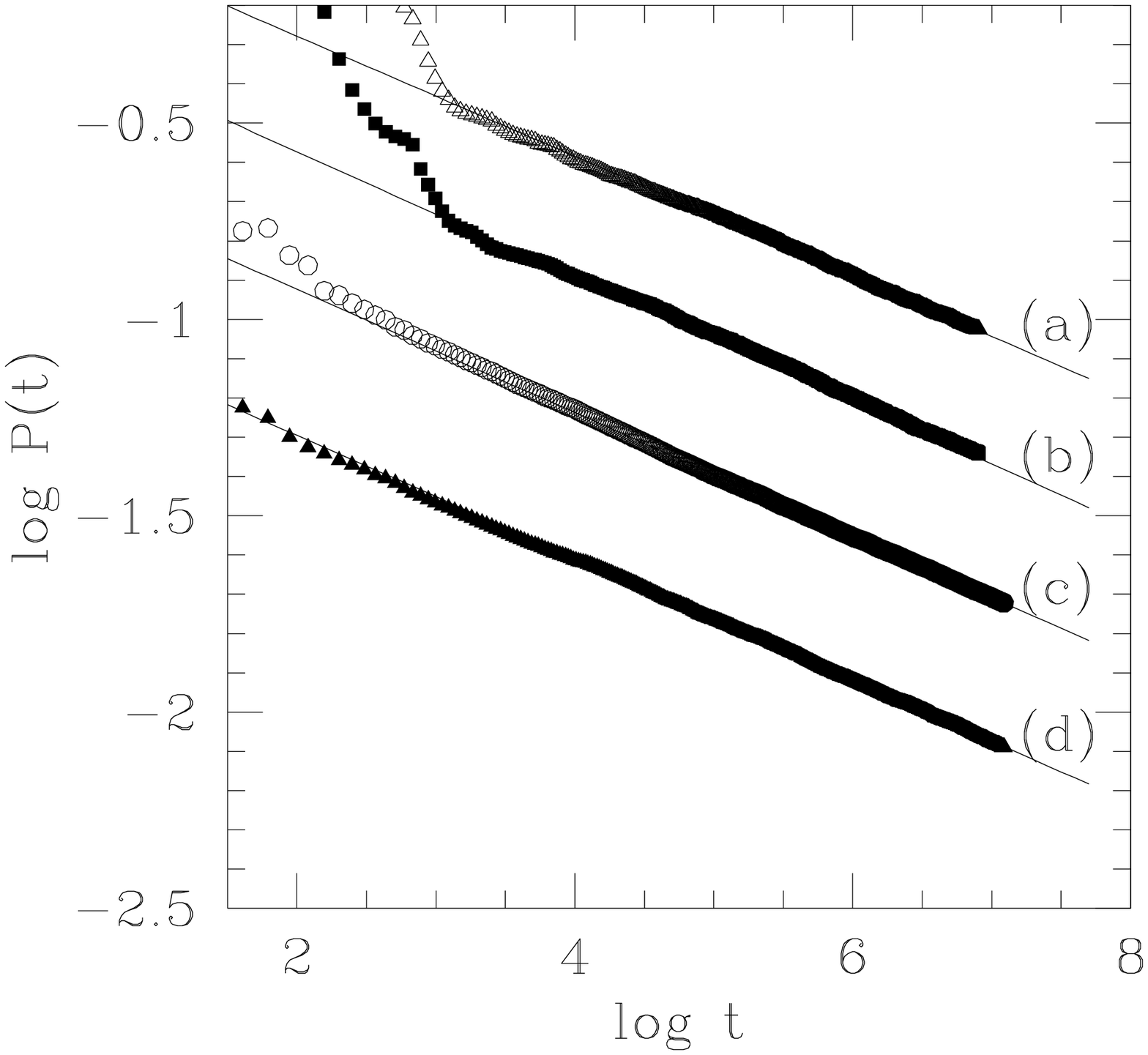,height=15truecm}}
\caption{ Log-log plot (base $e$) of of $P (t)$ vs $t$ for all 4 critical points:
  (a) $k = 1$, $\omega=0.068$, $\epsilon_c = 0.63775$; (b) $k = 1$,
  $\omega=0.064$, $\epsilon_c = 0.73277$; (c) $k = 3.1$,
  $\omega=0.18$, $\epsilon_c = 0.70100$ and (d) $k = 3.1$,
  $\omega=0.19$, $\epsilon_c = 0.65612$. Table II gives the exponent
  $\delta$ of the power law fits for the different critical points.}
\end{center}
\end{figure}

\begin{figure}[htb]
\begin{center}
\mbox{\psfig{file=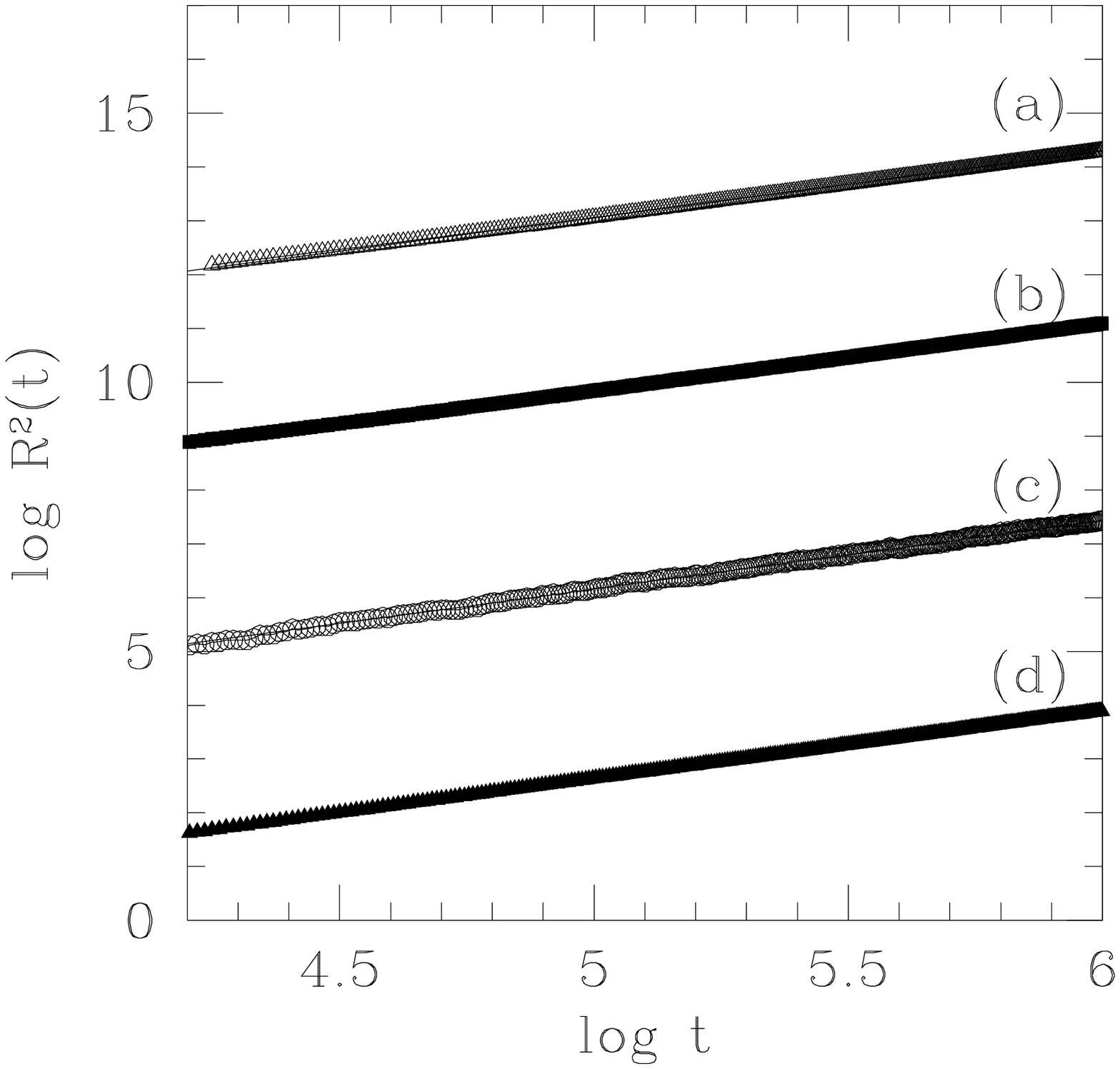,height=15truecm}}
\caption{ Log-log plot (base $e$) of $R^2(t)$ vs $t$ for all 4 critical points:
  (a) $k = 1$, $\omega=0.068$, $\epsilon_c = 0.63775$; (b) $k = 1$,
  $\omega=0.064$, $\epsilon_c = 0.73277$; (c) $k = 3.1$,
  $\omega=0.18$, $\epsilon_c = 0.70100$ and (d) $k = 3.1$,
  $\omega=0.19$, $\epsilon_c = 0.65612$. Table II gives the exponent
  $z_s$ of the power law fits for the different critical points.}
\end{center}
\end{figure}


\begin{thebibliography}{99}
\bibitem{type1} This type of STI is known as the type I STI in
literature.
\bibitem{singleseed} If we start with the very special initial
  background of all sites at some fixed $\theta_{bg}$, a single active
  seed will of course spread symmetrically. The exponents obtained
  from this symmetric spreading is non-universal and varies with
  $\theta_{bg}$. For instance if $\theta_{bg} = 0.4$ the spreading
  exponent $\eta = 0.07$, while if $\theta_{bg} = 0.0$ the $\eta =
  0.05$. However, introducing a very small spread around $\theta_{bg}$
  restores the DP universality.
\end{thebibliography}
\end{document}